\documentclass[%
 aip,
 groupedaddress,
 amsmath,amssymb,
 reprint,%
]{revtex4-1}

\usepackage{graphicx}
\usepackage{dcolumn}
\usepackage{bm}
\usepackage{color} 
\usepackage{CJK}

\def\la{{\langle}}
\def\ra{{\rangle}}

\newcommand{\beq}{\begin{equation}}
\newcommand{\eeq}{\end{equation}}
\newcommand{\beqa}{\begin{eqnarray}}
\newcommand{\eeqa}{\end{eqnarray}}

\begin{document}

\preprint{AIP/123-QED}

\title{Fast adiabatic-like spin manipulation in a two-electron double quantum dot}

\author{Yue Ban}

\address{Departamento de Qu\'{\i}mica-F\'{\i}sica, UPV/EHU, Apdo 644, 48080 Bilbao, Spain}

\date{\today}
\begin{abstract}
We apply the transitionless quantum driving method
to control the electron spin of a two-electron double quantum dot with spin-orbit coupling by time-dependent electric fields.
The $x$ and $y$ components of applied electric fields in each dot are designed to achieve fast adiabatic-like passage in the nanosecond timescale. To simplify the setup, we can further transform Hamiltonian by $z$ axis to design an alternative speed-up adiabatic passage, without using the applied electric field in $y$ direction.

\end{abstract}
\maketitle

\textit {Introduction}---
Spintronics aimes at fast and robust spin control in nanostructure \cite{spin resonance,Rashba,Nowack}.
There are several methods to manipulate spin accurately, such as electron spin resonance induced by magnetic
field oscillating at the Zeeman transition frequency \cite{spin resonance} and electric control with spin-orbit (SO) coupling \cite{Rashba}.
Recently proposed techniques of ``shortcut to adiabaticity" \cite{Chen10a,Berry09,Rice,Chen10b,ChenPRA} motivate us to achieve a high-fidelity control in a single quantum dot \cite{single-dot} (QD) and a two-electron double QD \cite{double-dot}. In a single QD, we applied inverse engineering method to design a fast and robust protocol of spin flip in the nanosecond timescale, based on
the Lewis-Riesenfeld theory \cite{single-dot}. Furthermore, in a two-electron QD, more freedoms of the applied electric fields provide the flexibility to
control the spin from one arbitrary state to the target state, by controllable Lewis-Riesenfeld phases. A different shortcut is provided by tansitionless quantum driving \cite{Berry09}, reformulated by Berry and equivalent to counter-diabatic control proposed by Demirplak and Rice \cite{Rice}. This technique was originally utilized to control the spin
in the fast adiabatic-like way by the applied magnetic fields \cite{Berry09,2spin-Bmethod}. Shortly afterwards, it has been extensively applied to various quantum systems
like two-level or three-level atoms \cite{Chen10b}, Bose-Einstein condensates in accelerated optical lattices \cite{Oliver} and
electron spin of a single nitrogen-vacancy center in diamond \cite{Suter}. 

In this Letter, we use transitionless quantum driving to design the external electric fields for rapid spin manipulation in a two-electron double QD in the presence of a static magnetic field and spin orbit coupling. Different from Berry's calculation \cite{Berry09}, we will apply the electric fields and take advantage of spin-orbit coupling, since the time-dependent electric fields are easy to be generated on the nanoscale by adding local electrodes \cite{Nowack}. In addition, in a single QD, there are only two controllable parameters, that is, $x$ and $y$ components of the electric fields, so that it is difficult or even impossible to produce the desired all-electrical interaction by transitionless quantum driving \cite{single-dot}. However, transitionless quantum driving is applicable in a two-electron double QD, as there exist more freedoms with four controllable parameters, $x$ and $y$ components of the external electric fields for each electron in double QD. To simplify the experimental setup and reduce the device-dependent noise, we can further apply the concept of multiple Schr\"{o}dinger pictures  \cite{Multiple-picture} to find an alternative shortcut with only $x$ component of the applied electric fields in our system, by $z$ rotation of Hamiltonian.

\begin{figure}[]
\begin{center}
\scalebox{0.5}[0.5]{\includegraphics{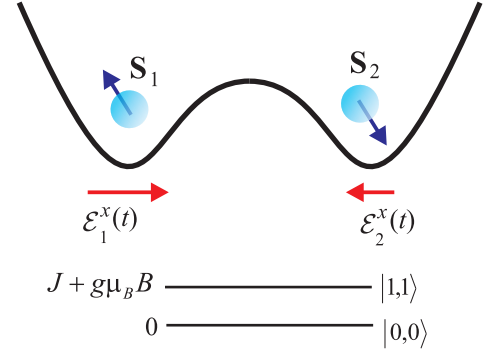}}
\caption{ (color online.) Schematic diagram of a two-electron double quantum dot in the presence of the external electric fields.}
\label{diagram}
\end{center}
\end{figure}

\textit{Hamiltonian}---
Two electrons are confined in a double QD, described as a quartic potential, where they are isolated by Coulomb blockade, illustrated in Fig. \ref{diagram}.
The spin-dependent Hamiltonian is $H_{\textrm{total}} = H_s + H_{\textrm{int}}$, where
\beqa
\label{H}
H_\textrm{s} = J \bm{S}_1 \cdot \bm{S}_2 + \sum_j \frac{\Delta_{j}}{2} \sigma^z_{j}.
\eeqa
Here $j=1,2$ represent the electrons $1$ and the electron $2$. $\bm{S}_j = \bm{\sigma}_j/2$, are the spin operators of two electrons. The Zeeman term $\Delta_{j} = g \mu_B B_j$, where $\mu_B=\hbar|e|/(2 m_0 c)$ is the Bohr magneton, $g$ is the Land\'{e} factor with negative value ($g <0$), like in GaAs and InAs, and $m_0$ is the mass of the free electron. $B_1$ and $B_2$ are the static magnetic fields applied to the electron $1$ and $2$ in $z$ direction, respectively. For simplicity, we set $B_1 = B_2 = B$, so that $\Delta=\Delta_1=\Delta_2 = g \mu_B B$. In the presence of the applied magnetic field, the lowest four eigenstates of the system can be expressed by singlet and triplet for $S=0$ and $S=1$ in the basis of $|S,S_z\ra$. If the energy difference between the singlet $|0,0 \ra$ and the lowest one of the triplet $|1,1 \ra$ is much less than the gap between the singlet and the triplet $J$, which means
$|J + \Delta| \ll J $,
we focus on the state transition between these lowest two states $|0,0\ra$ and $|1,1\ra$, as shown in Fig. \ref{diagram}.
By choosing $ |0,0\ra = (1, 0)^T $ and $|1,1\ra  = (0, 1)^T $, we can write the reduced Hamiltonian $H_s$ in the form of $2 \times 2$ matrix,
\beqa
\label{H1}
H_\textrm{s} = \left(
\begin{array}{cc}
-\displaystyle{\frac{3}{4}} J & 0
\\
0 & \displaystyle{\frac{1}{4}} J + \Delta
\end{array}
\right).
\eeqa

The interactions between the electric field and the electron  $H_{\textrm{int}}$ are expressed as,
\beqa
\label{Hint1}
H_{\textrm{int}} = -\frac{e}{c} \sum_j \textbf{A}_j \cdot \textbf{v}_j,
\eeqa
where the vector potential $\textbf{A}_j(t)$  are related to the external electric fields, ${\bf \mathcal{E}}_1 (t) = -(1/c)\partial \textbf{A}_1/\partial t$ and ${\bf \mathcal{E}}_2 (t) = -(1/c)\partial \textbf{A}_2/\partial t$, and $\textbf{v}_j$ are the spin-dependent velocity operators. We consider the SO coupling including structure-related Rashba ($\alpha$) term and bulk-originated Dresselhaus ($\beta$) term for $[1 1 0]$ growth axis,
\begin{eqnarray}
\label{Hsoc}
H_{\textrm{soc}} = \sum_j \alpha (\sigma^x_j p^y_j - \sigma^y_j p^x_j) + \sum_j \beta \sigma^z_j p^x_j,
\end{eqnarray}
so that the spin-dependent velocity operators become
\beqa
v^x_j  &=& \frac{i}{\hbar} \left[H_{\rm soc}, x_j\right] = \beta \sigma^z_j - \alpha \sigma^y_j,
\\
v^y_j  &=& \frac{i}{\hbar} \left[H_{\rm soc}, y_j\right] = \alpha \sigma^x_j.
\eeqa
Therefore, the total spin-dependent Hamiltonian $H_{\textrm{total}}$ is
\beqa
\label{Htotal}
H_{\textrm{total}} = \frac{\hbar}{2} \left(
\begin{array}{cc}
Z_1 & X + i Y
\\ X - i Y & Z_2
\end{array}
\right).
\eeqa
where
\beqa
\label{X_H}
X &=& \frac{\sqrt{2}\alpha}{\hbar} \frac{e}{c} (A^y_{1} - A^y_{2}),
\\
\label{Y_H}
Y &=& -\frac{\sqrt{2}\alpha}{\hbar} \frac{e}{c} (A^x_{1} - A^x_{2}),
\\
\label{Z1_H}
Z_1 &=&  -\frac{3J}{2 \hbar},
\\
\label{Z2_H}
Z_2 &=& \frac{1}{\hbar}\left[ \frac{J}{2} + 2 \Delta - 2 \frac{e}{c} \beta (A^x_{1} + A^x_{2})\right]. 
\eeqa
%
To rewrite the symmetric Hamiltonian, we shift one quantity
%
$Z_0 = -J/4 + \Delta/2 - (e\beta/2c) (A^x_1 + A^x_2)$,
%
and finally obtain Hamiltonian $H_{\textrm{total}} = Z_0 \hat{I} + H$,
where
%
%
\beqa
\label{Htotal}
H = \frac{\hbar}{2} \left(
\begin{array}{cc}
Z & X + i Y
\\ X - i Y & -Z
\end{array}
\right),
\eeqa
with
%
$Z = (1/\hbar) \left[-J - \Delta + (e\beta/c)(A^x_1 + A^x_2)\right]$.
%
The solution to the Schr\"{o}dinger equation of $H$ differentiates from that of $H_{\textrm{total}}$ by 
the factor $\exp[-i\int Z_0(t')d t']$, while the energy level $|0,0\ra$ and $|1,1\ra$ are shifted by $Z_0$.
The states after the shifting 
are denoted by $|1\ra$ and $|-1\ra$, respectively, and their populations
remain unchanged as the ones of the previous states, $|0,0\ra$ and $|1,1\ra$.

\textit{Transitionless fast spin tranfer}---
Our aim is to transfer the spin from $|-1 \ra$ to $|1\ra$ totally during a reasonably short time duration. The form of Hamiltonian $H$ in Eq. \ref{Htotal} tells us that $Y$ and $Z$ are the functions of $A^x_1$ and $A^x_2$ and $X$ is the function of $A^y_1$ and $A^y_2$. Different from the Hamiltonian of one electron confined in a single dot \cite{single-dot}, the transitionless quantum driving can be applicable to the spin control in a two-electron double QD, as we can figure out how the Hamiltonian $H$ (including reference Hamiltonian $H_0$ and counter-diabatic term $H_1$) is implemented by corresponding electric fields. We may take the reference Hamiltonian $H_0$ as
\beqa
\label{H0}
H_0 = \frac{\hbar}{2} \left(
\begin{array}{cc}
Z & i Y
\\ - i Y & -Z
\end{array}
\right),
\eeqa
driven by $A^x_1$ and $A^x_2$. The example of a double QD of GaAs-based structure is considered below, where $g=-0.44$ and the static magnetic fields are $B_1=B_2=3.43$ T. The energy gap between the singlet and the triplet is $J = 0.1$ meV, so that $|J + \Delta| / J = 0.12 \ll 1$ with the above parameters.

\begin{figure}[]
\begin{center}
\scalebox{0.6}[0.6]{\includegraphics{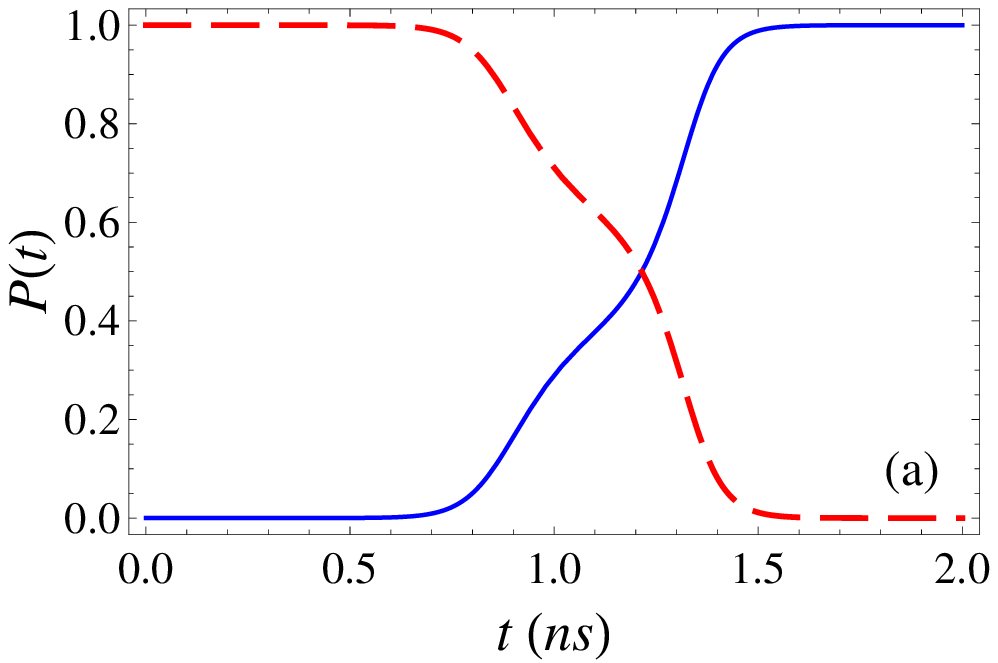}}
\scalebox{0.6}[0.6]{\includegraphics{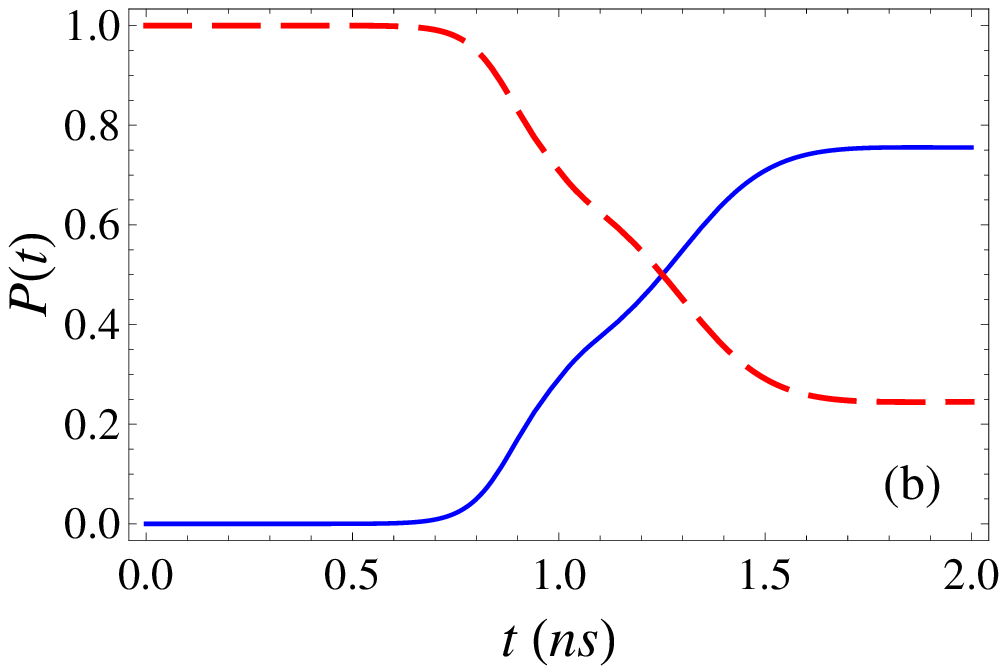}}
\scalebox{0.6}[0.6]{\includegraphics{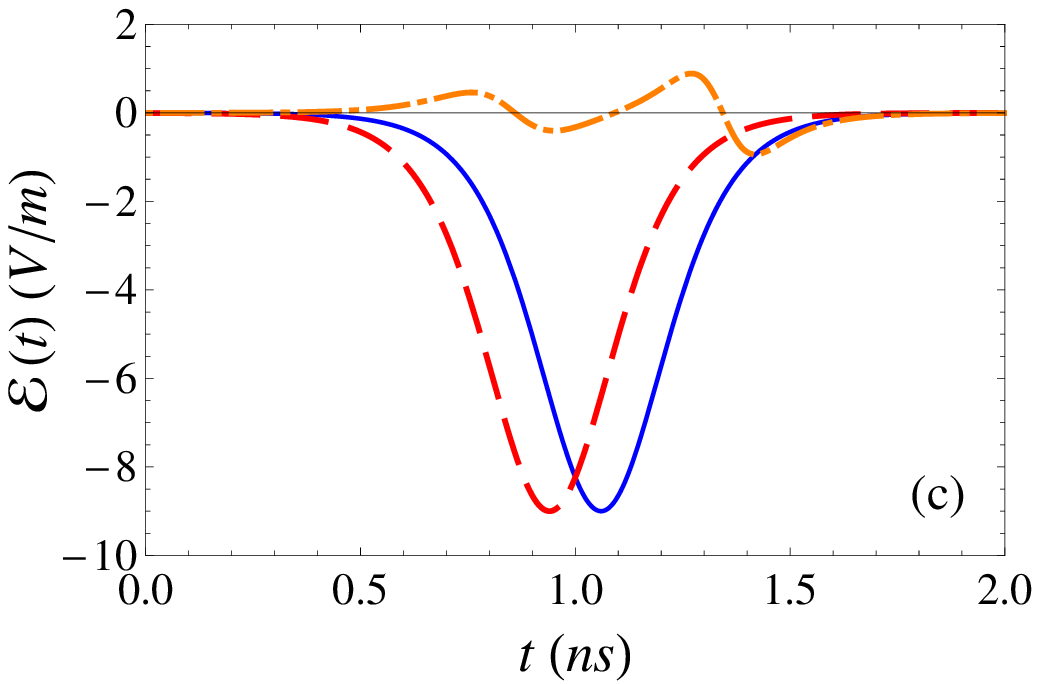}}
\caption{ (color online.) (a) Time evolution of the populations $P_1^{\textrm{in}}(t)$ (solid blue line) and $P_{-1}^{\textrm{in}}(t)$ (dashed red line) as the instantaneous eigenstates of $H_0$,
which coincide with the populations $P_1(t)$ and $P_{-1}(t)$ as the solution to the Schr\"{o}dinger equation of $H$.
(b) Time evolution of the populations $P_1^0(t)$ (solid blue line) and $P_{-1}^0(t)$ (dashed red line) as the solution to the Schr\"{o}dinger equation of $H_0$, showing that this is not an adiabatic process.
(c) The applied electric fields in the $x$ direction ${\mathcal E}^x_1$ (solid blue line) and ${\mathcal E}^x_2$ (dashed red line), and the additional two electric fields in the $y$ direction with the difference ${\mathcal E}^y_D = {\mathcal E}^y_1-{\mathcal E}^y_2$ (dot-dashed orange line) drive the population inversion. Other Parameters: $t_f=2$ ns, $\hbar \alpha = 1.2 \times 10^{-6}$ meV $\cdot$ cm, $\hbar \beta = 0.3 \times 10^{-6}$ meV $\cdot$ cm.} \label{P-transfer}
\end{center}
\end{figure}

With the help of reference Hamiltonian (\ref{H0}), we can write down the instantaneous eigenstates, $|\chi_\pm \ra$, satisfying $H_0 |\chi_\pm \ra = E_\pm |\chi_\pm \ra$,
where the instantaneous eigenvalues are $E_\pm = \pm \hbar\sqrt{Z^2+Y^2}/2$, and the instantaneous eigenstates are
\beqa
|\chi_{+} \ra
=\left(\begin{array}{c}
\cos\displaystyle{\frac{\theta}{2}}  e^{i \varphi}
\\
\sin\displaystyle{\frac{\theta}{2}}
\end{array}
\right),
|\chi_{-} \ra
=
\left(\begin{array}{c}
 \sin \displaystyle{\frac{\theta}{2}}
\\
- \cos \displaystyle{\frac{\theta}{2}} e^{-i \varphi}
\end{array}
\right),~~~~~
\eeqa
with the mixing angle $\theta=\arccos[Z/(Y^2+Z^2)]$ and $\varphi=\pi/2$. Once the adiabaticity condition \cite{Berry09,ChenPRA}
\beqa
\label{adiabatic-condition}
\left|\frac{Z \dot{Y} - Y \dot{Z}}{(Y^2 + Z^2)^\frac{3}{2}}\right| \ll 1
\eeqa
is fulfilled, the state $|\Psi^0 \ra$, the solution to the Schr\"{o}dinger equation of $H_0$, evolves from $|\Psi^0(0) \ra = |\chi_\pm(0) \ra$ and follows the adiabatic approximation
\beqa
\label{adiabatic-approximation-wavefunction}
|\Psi^0(t)\ra = \exp \left[ -\frac{i}{\hbar} \int^t_0 d t' E_\pm(t') \right] |\chi_\pm(t) \ra.
\eeqa
Otherwise, transitions between $|\chi_\pm(0) \ra$ will occur. To implement population inversion, from $|-1\ra$ to $|1\ra$, along one of instantaneous eigenstate, $|\chi_{+} (t) \ra$, we set the ansatz of the vector potential $A^x_{j} = A_0 \tanh[(t - a_j t_f)/(w_j t_f)]$, where $a_j$, $w_j$ describe the change rate of $A^x_{j}$ and $j=1,2$. 
To fulfill the initial and final states, $Y(0)=Y(t_f)=0$ should be fixed, which means at the initial and final times $A^x_1$ and $A^x_2$ are equal to each other. Meanwhile, the mixing angle $\theta$ goes from $\pi$ to $0$, crossing the point $\pi/2$ during the interval $(0,t_f)$. With this strategy, we produce the reference electric fields $\mathcal{E}^x_j$, displayed in Fig. \ref{P-transfer} (c). For the following comparison, we first show the dynamics of populations for the instantaneous eigenstates, $P^{\textrm{in}}_{1} (t) = |\la 1 | \chi_+ (t)\ra|^2$ and $P^{\textrm{in}}_{-1} (t)= |\la -1 | \chi_+ (t)\ra|^2 $ (seen in Fig. \ref{P-transfer}(a)).
In practice, this process is not adiabatic, so the populations of exact solution, $\Psi^0 (t)$, of Hamiltonian $H_0$ are obtained as $P_1^0(t_f)= |\la 1 | \Psi^0 (t_f)\ra|^2 =0.76$ and $P_{-1}^0(t_f)= |\la -1 | \Psi^0 (t_f) \ra|^2 =0.24$, (seen in Fig. \ref{P-transfer}(b)), which are not consistent with that of the instantaneous eigenstates. Of course, the adiabatic passage can be realized by extending $t_f$ and increasing the electric fields, respectively. For example, if we prolong $t_f=14$ ns and keep the previous $A^x_j$, the process will become adiabatic, and $P_1^0(t_f) = 0.9999$ is finally achieved. On the other hand, $P_1^0(t_f)=0.9999$ can be also achieved, when the magnitude of ${\mathcal E^x_j}$ are increased by $11.5$ V/ m and keep $t_f=2$ ns.

Next, transitionless quantum driving will provide supplementary time-dependent interactions $H_1$ that cancel the diabatic couplings of a reference process $H_0$, and make the reference
process fast and adiabatic-like. The supplementary counter-diabatic term $H_1 = \sum_\pm|\partial_t \chi_\pm \ra \la\chi_\pm|$ is \cite{Berry09,Chen10b}
\beqa
\label{H1}
H_1 = \frac{\hbar}{2} \left(
\begin{array}{cc}
0 & X
\\ X & 0
\end{array}
\right),
\eeqa
driven by $A^y_1$ and $A^y_2$, where $X=\dot\theta=(\dot{Y} Z-Y \dot{Z})/(Z^2 + Y^2)$. As a result, the solution $\Psi(t)$ to the Schr\"{o}dinger equation of $H = H_0 + H_1$ becomes exactly the  adiabatic approximation of $H_0$. The corresponding dynamics of the populations, $ P_1(t)= |\la 1 | \Psi(t)\ra|^2$ and $ P_{-1}(t)= |\la -1 | \Psi(t) \ra|^2$. The populations $P_1(t)$ and $P_2(t)$ coincide with $P_1^{\textrm{in}}(t)$ and $P_2^{\textrm{in}}(t)$ respectively, as shown in Fig. \ref{P-transfer}(a). The difference between two additional $y$ components is $A^y_D = A^y_1 - A^y_2 = \hbar \dot\theta /(\sqrt{2} e \alpha)$ and the corresponding time-dependent function of ${\mathcal E}^y_D = {\mathcal E}^y_1 - {\mathcal E}^y_2$ is plotted in Fig. \ref{P-transfer}(c). The maximal magnitudes of ${\mathcal E}^y_D$ is $0.94$ V/m. Obviously, they are much less than the increasing values in magnitude of ${\mathcal E}^x_j$ to achieve the adiabatic process, as mentioned above.
This implies that the transitionless quantum driving can really speed up the adiabatic process.
As a matter of fact, ${\mathcal E}^y_D$, as the function of $\dot\theta$, is related to $\dot{Y}$ and $\dot{Z}$.
The shorter time is, the larger value of ${\mathcal E}^y_D$ is required. To implement ${\mathcal E}^y_D$ easily in the experiment,
we need the smooth function of ${\mathcal E}^y_D$, therefore in general ${\mathcal E}^x_1$ and ${\mathcal E}^x_2$ should not vary very dramatically.

\textit{$z$-axis rotation}---
In reality, the electron spin is subject to the device-dependent noise, which could be the amplitude noise of the electric fields \cite{single-dot}. It can be quite important, especially when the electric fields are relatively weak. From the above analysis, we find that four controllable parameters, ${\mathcal E}^x_{j}$ and ${\mathcal E}^y_{j}$, $x$ and $y$ components of the electric fields for each electron in double QD should be applied. If $y$ component of the electric fields can be reduced,  we can decrease decoherent effects resulting from the device-dependent noise. To this end, we can apply the concept of multiple Schr\"{o}dinger pictures, and make unitary transformation of Hamiltonian $H$ by $z$-axis rotation \cite{Multiple-picture}.
We write down the dynamical Hamiltonian as follows
\beqa
\label{Hs}
H = \frac{\hbar}{2} \left(
\begin{array}{cc}
Z & i Q e^{i(\phi-\pi/2)}
\\-i Q e^{-i(\phi-\pi/2)} & -Z
\end{array}
\right),
\eeqa
where $\tan \phi = Y / X$ and $Q = \sqrt{X^2+Y^2}$.
\begin{figure}[] \begin{center}
\scalebox{0.6}[0.6]{\includegraphics{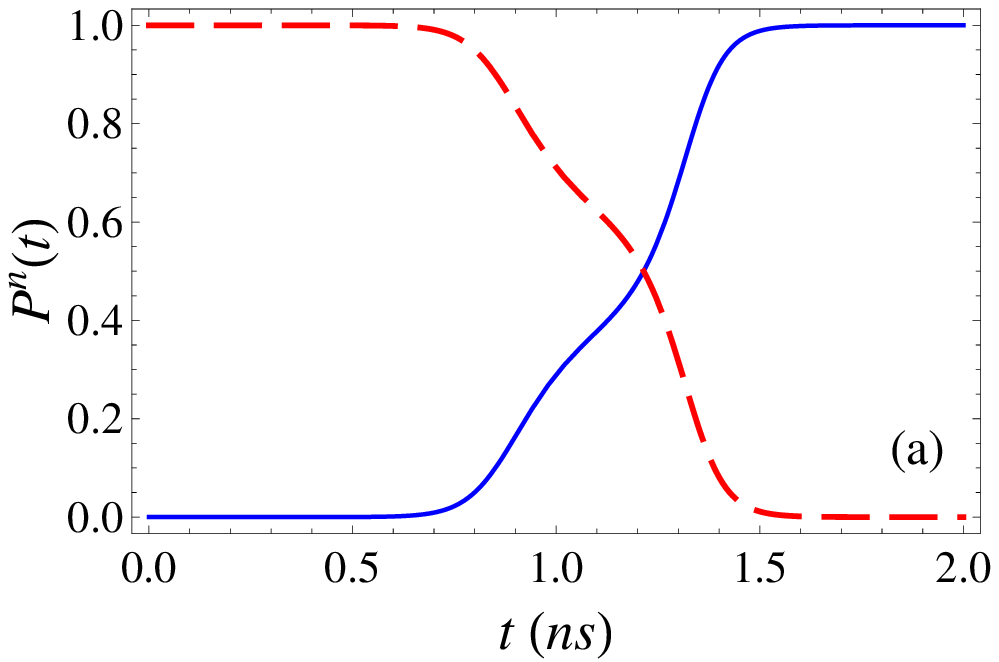}}
\scalebox{0.6}[0.6]{\includegraphics{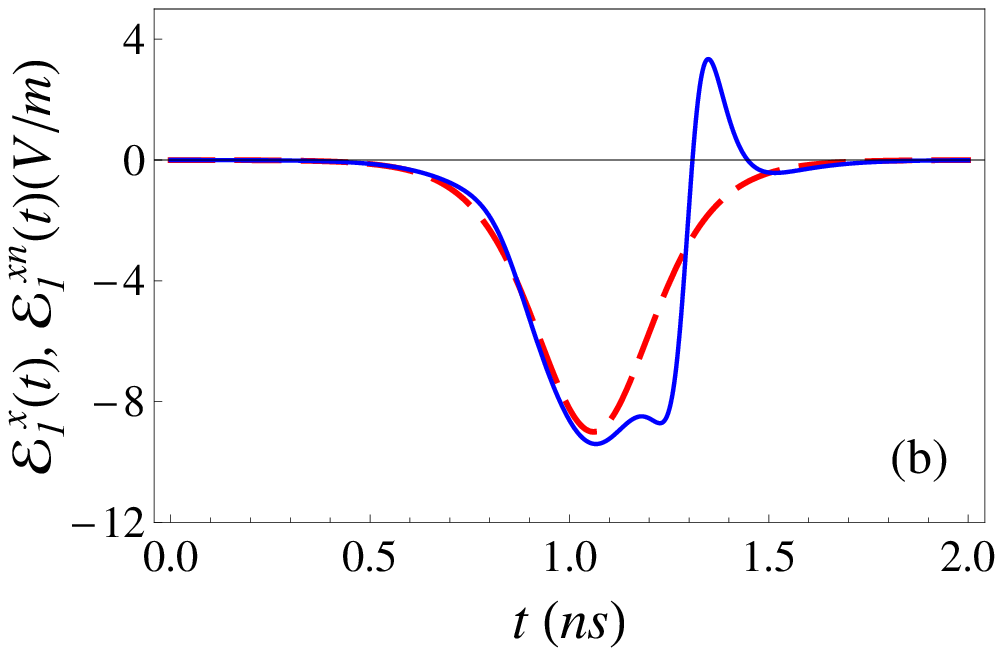}}
\scalebox{0.6}[0.6]{\includegraphics{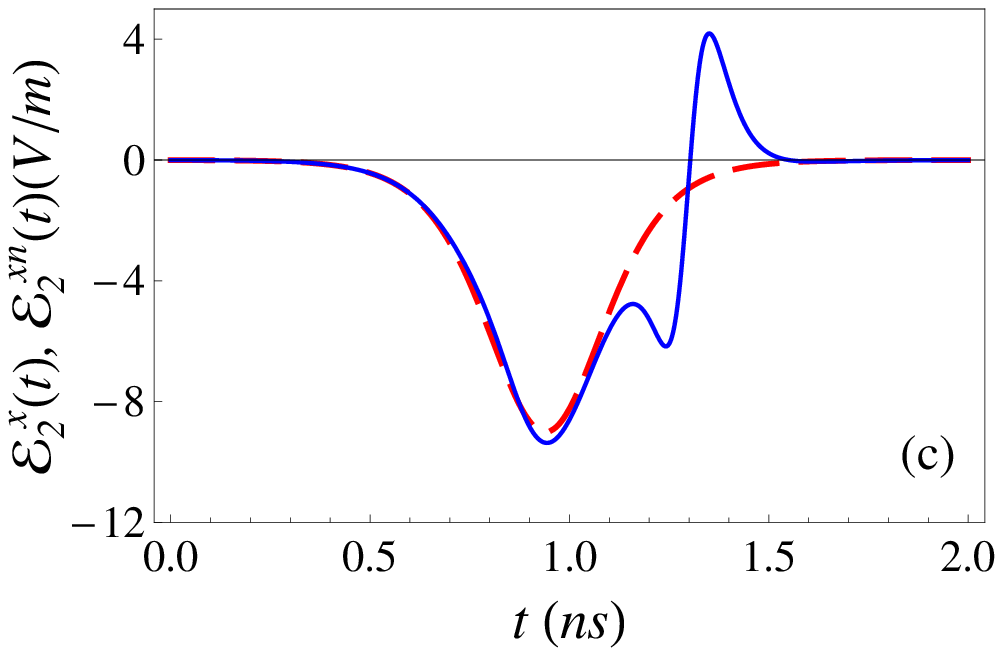}}
\caption{
(color online.) (a) The populations $P^n_1(t)$ and $P^n_{-1}(t)$ as the solution to the Schr\"{o}dinger equation $H'$. 
(b) Comparisons between ${\mathcal E}^{xn}_1$ (solid blue line) and ${\mathcal E}^{x}_1$ (dashed red line). (c) Comparisons between ${\mathcal E}^{xn}_2$ (solid blue line) and ${\mathcal E}^{x}_2$ (dashed red line).
Other parameters are the same as in Fig. \ref{P-transfer}. } \label{PI-correction}
\end{center}
\end{figure}
By applying the unitary transformation \cite{Berry90}
\beqa
\label{U}
U = \frac{\hbar}{2} \left(
\begin{array}{cc}
e^{i(\phi-\pi/2)} & 0
\\ 0 & e^{-i(\phi-\pi/2)}
\end{array}
\right),
\eeqa
which amounts to a rotation around $z$ axis by the angle $\pi/2-\phi$,
we calculate the new Hamiltonian $H' = U^\dag(H-K)U$ with $K = i\hbar \dot U U^\dag$, and finally obtain
\beqa
\label{Hs'}
H' = \frac{\hbar}{2} \left(
\begin{array}{cc}
Z+\dot\phi & i Q
 \\ -i Q & -Z-\dot\phi
\end{array}
\right),
\eeqa
%
without $\sigma_x$ term.
%
%
We should notice that the dynamics of Hamiltonian $H =H_0 +H_1$ and $H'$ is not the same (the populations are the same because of $z$ rotation). However, the Hamiltonian $H'$ is equal to the original one $H =H_0 +H_1$ at $t=0$ and $t_f$, which guarantees that the initial (final) states of $H$ and $H'$ coincide. So the Hamiltonian $H'$ can provide an alternative way to implement the shortcuts to adiabaticity.
According to the Hamiltonian $H'$ Eq. (\ref{Hs'}), we may acquire two new controllable parameters, $\mathcal{E}^{xn}_1$ and $\mathcal{E}^{xn}_2$, $x$ component of the electric fields, since $Z+\dot\phi$ and $Q$ are the functions of the sum and the difference of $A^{xn}_1$ and $A^{xn}_2$, respectively.
The solution, $\Psi^n (t)$, of the Schr\"{o}dinger equation of $H'$ can be solved numerically, and the populations $P^n_1(t)=|\la 1 | \Psi^n (t)\ra|^2$ and $P^n_{-1}(t)=|\la -1 | \Psi^n (t) \ra|^2$, are shown in Fig. \ref{PI-correction} (a). At the final time, $P_1^n(t_f)=1$ and the population is completely inverted. The new electric fields only in $x$ direction are shown in Fig. \ref{PI-correction} (b-c) with some corrections compared with the previous ones ${\mathcal E}^x_1$ and ${\mathcal E}^x_2$.

\textit{Conclusion}---
We propose the shortcuts to manipulate the spin states formed in a two-election double QD by using transitionless quantum driving. The Hamiltonian $H$ is divided into two parts, the reference process $H_0$, driven by $A^x_1$ and $A^x_2$, and the supplementary time-dependent interaction $H_1$, driven by $A^y_1$ and $A^y_2$. By applying $x$ and $y$ components of electric fields for each electron, the spin system follows exactly the adiabatic approximation of the reference Hamiltonian $H_0$, in the time scale of nanosecond. In order to simplify the setup, and decrease the device-dependent noise effect, we further transform the Hamiltonian by $z$ axis and obtain the new Hamiltonian implemented only by $x$ component of electric fields.
This provides an alternative shortcut to realize the fast and adiabatic-like spin control. We hope these results may lead to the applications
in spintronics and quantum information processing with the state-of-the-art technique.

\textit{Acknowledgement}---
Y. B. acknowledges financial support from the Basque Government (Grant Nos. BFI-2010-255 and IT472-10),
Ministerio de Ciencia e Innovacion (Grant No. FIS2009-12773-C02-01), and the UPV/EHU under program UFI 11/55.
Valuable discussions from E. Ya. Sherman and X. Chen are appreciated.

\end{document}